# Nanorods of Well-Defined Length and Monodisperse Cross-Section Obtained from Electrostatic Complexation of Nanoparticles with a Semiflexible Biopolymer

*Li Shi,[1,2] Florent Carn,[1,*] François Boué,[2] Gervaise Mosser,[3] and Eric Buhler[1,2,*]*

[*]     Dr. Florent Carn, Prof. Eric Buhler
[1] Laboratoire Matière et Systèmes Complexes (MSC), UMR 7057, Université Paris Diderot-Paris 7, Bâtiment Condorcet, 75205 Paris cedex 13, France
E-mail: eric.buhler@univ-paris-diderot.fr
        Li Shi, Dr. Florent Carn, Prof. Eric Buhler
[1] Laboratoire Matière et Systèmes Complexes (MSC), UMR 7057, Université Paris Diderot-Paris 7, Bâtiment Condorcet, 75205 Paris cedex 13, France
        Li Shi, Prof. François Boué, Prof. Eric Buhler
[2] Laboratoire Léon Brillouin, UMR 12 CEA-CNRS, CEA Saclay, 91191 Gif-sur-Yvette, France
        Dr. Gervaise Mosser
[3] Laboratoire Chimie de la Matière Condensée de Paris, UMR 7574, UPMC, Collège de France, 11 place Marcelin Berthelot, 75005 Paris, France

Nanometric rods technological applications are many but a severe restriction is the complexity of most syntheses. The process we present is surprisingly simple, cheap and makes use of an abundant natural polymer. Nanometric rods open routes toward new applications such as plasmon based wave guide, biosensors, nanorulers or theragnostic materials.[1] Of course, they critically depend on the controlled ordering of the nano-building block that determines their collective interactions.[2] Opposite to elaborated procedures such as for example, rodlike nanocrystals or well controlled 1D structures built on a surface, is the more basic approach of 1D assembly of preformed spherical nanoparticles (NPs), now largely available. This can be template or non-template assisted assembly.[3] In the first case, rigid polymers,[4] viruses, or carbon nanotubes[5] have served as template to assemble NPs.[1b,1c,1j,1k,4-8] In the second case, the anisotropic or isotropic functionalization of nanoparticles by electrolytes, surfactants or biomolecular ligands has already enabled the directional assembly of nanoparticles.[1e,1f,9-13] Still the problem, generally, is the involvement of multi-step protocols that necessitate sophisticated chemical and/or physical methods. Moreover, this complexity combines with the one of a "shape diagram", since these processes can also lead to either extended ramified chains or very small entities (dimers or trimers) instead of rods with finite and intermediate size.[9]

Our simple alternative is based on electrostatic complexation between spherical inorganic nanoparticles (widely known), and an abundant natural semi-flexible polyelectrolyte of opposite charge. Electrostatic associations, which can lead to solid-liquid or liquid-liquid separation, have attracted a considerable theoretical and experimental attention in the last two decades. A great variety of nanostructured complexes has been reported but to the best of our knowledge, significant control over morphology is scarcely claimed, apart from the preparation of isotropic shapes or necklace-like aggregates-a few of them being characterized in great detail.[14,15] Although post-treatments for stretching preformed physical complexes into highly ordered 1D assembly have been reported,[16] to date one pot preparations of nanometric rod shaped complexes are almost nonexistent.

Using this electrostatic complexation approach, after simple mixing of model Silica NPs (SiNPs) and cationic chitosan chains solutions, we obtain for the first time a stable dispersion of well-defined nanometric rods. The evidencing technique is Small Angle X-rays Scattering (SAXS), which enables us to estimate the characteristic sizes and the local structural parameters of the rods: (i) the number per rod of NPs (10 in average) and chitosan chains (1 to 2); (ii) the rod length (~200 nm) and its axial radius $R_{rod}$, equal to the one of the SiNPs ($R_{SiNP}$ ~ 10 nm). It is interesting to compare the NP radius with the polymer chain intrinsic persistence length $Lp \approx 7.5$ nm.[17] We see that $Lp/R_{SiNP} \approx 1$ and can then propose a mechanism of formation, where chitosan chains wrap around the particles. To compare to an illustrious case, let us note that this is in striking contrast with DNA packaging into chromatin, where an almost rigid polymer with larger persistence length $Lp$ ~ 50 nm is compacted around small oppositely charged histone with $R \approx 3.5$ nm. As a complementary technique, in real space, we used cryogenic transmission electronic microscopy (cryo-TEM); combined with SAXS this gives a wide range of observation scale from the nanometer to the micron scale.

First the two genuine components had to be carefully characterized. The stock solutions of chitosan or SiNPs suspensions were previously prepared (at twice the final concentration before mixing) in water buffer with $C_{CH3COOH} = 0.3$ M and $C_{CH3COONa} = 0.2$ M. Both systems are charged: about SiNPS, in these conditions, we have roughly estimated from the electrophoretic mobility ($\mu=-2.3.10^{-8}$ m².V$^{-1}$.s$^{-1}$) that each SiNP display 25 elementary negative charges via the Hückel equation. The autocorrelation function of concentration fluctuations, $g^{(1)}(q,t)$, measured by dynamic light scattering (DLS), decreases monoexponentially with a characteristic relaxation time inversely proportional to $q^2$ and leads to $R_{H,SiNP}$=12 nm. About chitosan, we briefly recall that it is a linear cationic polysaccharide derived by alkaline deacetylation of chitin in crustacean shells and is constituted of two residues linked by β-(1→4) glycosidic bonds, N-acetyl glucosamine and glucosamine (deacetylated fraction measured by NMR [17] $f_D$=87.5%), which is charged. In the preparation conditions, it exhibits a high polyelectrolyte character with approximately one positive charge per segment (of size 5 Å). The weight-average molecular weight of chains, $M_W$=313K (~2000 segments), and the radius of gyration, $R_G$=66 nm, were deduced from static light scattering (SLS) measurements and a classical Guinier analysis, and $R_{H,chitosan}$=44 nm (hydrodynamic radius) from DLS. Thus in the dilute range of chitosan concentration ($C_{chitosan}$ =0.1 g/l), chitosan chains are in the typical semi-rigid polymer-like conformation with $R_G/R_H$=66/44~1.5. In summary, both chitosan chains and SiNPs are well dispersed and stable in solution.

**Figure 1** illustrates the macroscopic phase behavior of the mixtures determined by visual inspection at 20 °C for molar ratio, r = $C_{chitosan}$ / $C_{SiNPs}$, comprises between r = 8.6 and r

= 0.0013 with constant $C_{chitosan}$ = 0.01 g/l. The samples are obtained after continuous stirring for several hours and let aging 2 weeks in static condition; two phase boundaries are detected at r* = 1.88 ± 0.20 and r** = 0.030 ± 0.005. In domains I and III of the phase diagram, obtained for r < r* and r > r** respectively, stable transparent single phase samples were observed. In the intermediate domain n°II, turbid samples progressively evolve under aging toward macroscopically biphasic samples ; both the supernatant and the dense lower phase display a liquid character according to qualitative flowing tests. Besides, it was also found that the phase boundaries of the system vary with either pH or ionic strength. This kind of liquid–liquid phase separations have often been observed in mixed systems involving polyelectrolytes and oppositely charged colloids and were generally discussed in the frame of associative phase separation or complex coacervation processes.[14,15] In the frame of this letter, we propose a thorough structural study of the soluble complexes obtained in monophasic domain n°III.

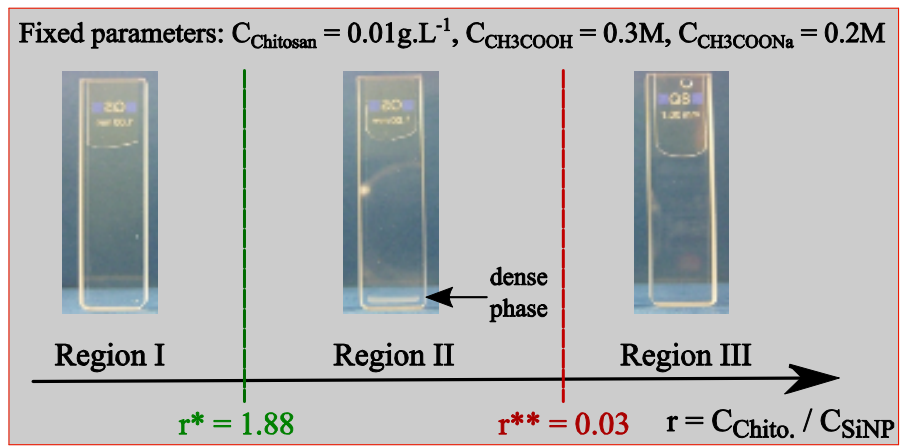

Figure 1. Schematic phase diagram of chitosan/SiNP mixtures at $C_{chitosan}$=0.01g/l. Along the horizontal axis are reported the different concentration thresholds (r* (region I/II) and r** (region II/III)) observed in the presence of $C_{CH3COOH}$ = 0.3 M and $C_{CH3COONa}$ = 0.2 M at 20°C.

SAXS experiments were carried out at T=20°C on the instrument ID-02 (ESRF, Grenoble) with configurations allowing a large $q$ range varying between 0.0011 Å$^{-1}$ and 0.57 Å$^{-1}$. The final spectra are given in absolute units of cross-section (cm$^{-1}$) following the standard procedures.[18] For the individual SiNPS, **Figure 2a** (lower curve) shows, in agreement with light scattering, that NPs are dispersed individually. As detailed in the Experimental Section, the form factor oscillations, damped by a size distribution, are well reproduced with $I(q)$ calculated as indicated in the Experimental Section. The SiNPs solution is well represented by a suspension of hard spheres with $R$=9.2 nm, and a variance σ=0.12. Extrapolation of the scattered intensity to zero-wave vector, $I(0)$, gives the weight-average molecular weight, $M_{W,SiNP}$=3×10$^6$ g/mol.

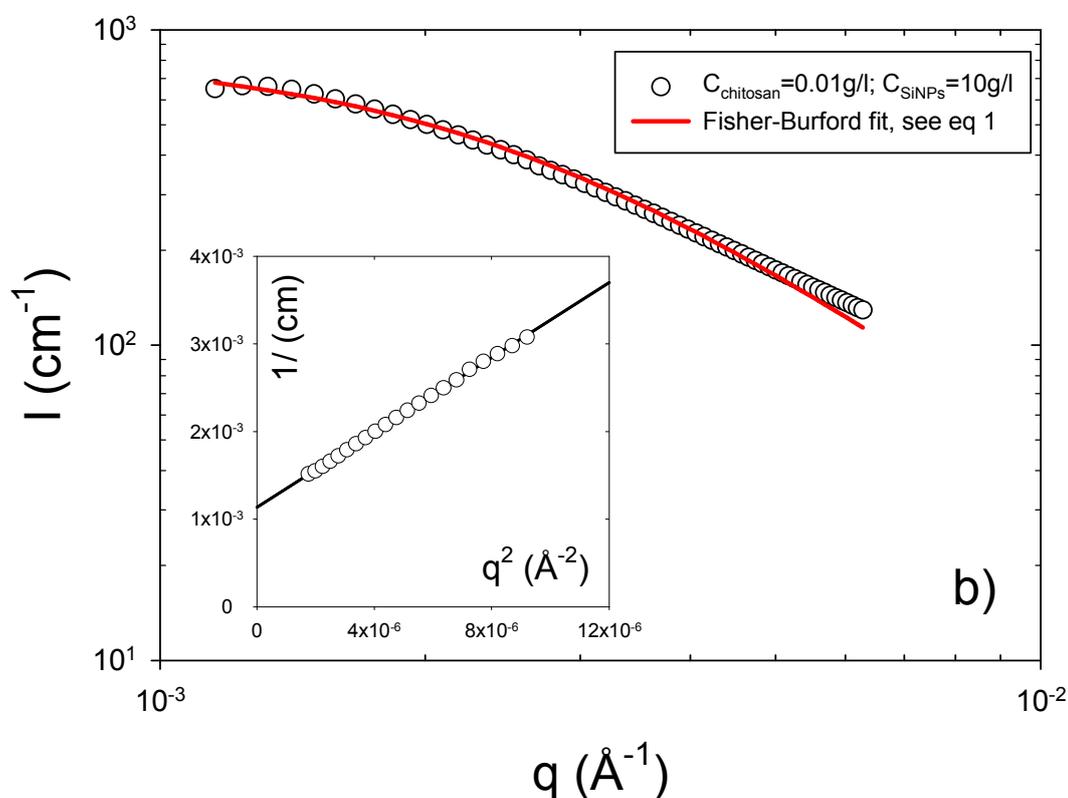

Figure 2. (a) SAXS spectra obtained for a $C_{chitosan}$=0.01g/l / $C_{SiNP}$=10g/l solution (upper curve) and for a $C_{SiNP}$=5g/l solution (lower curve) with $C_{CH3COOH}$ = 0.3 M and $C_{CH3COONa}$

= 0.2 M. For clarity the curves have been shifted by two log units along the y-axis with respect to each other. The continuous lines represent the fits of the data. (b) Low-$q$ scattering curves in a log-log representation. The solid line represents the Fisher-Burford fit with $D_f$=1, whereas the inset represents $I(q)^{-1}$ versus $q^2$ and the best linear fit.

For complexes (upper curve), due to the high electronic density and concentration of the SiNPs, the signal is dominated by the scattering of the SiNPs (whether they belong to complexes or not). Due to the small concentration and molecular weight of chitosan chains their signal is negligible. The scattering curve exhibits the overall behaviour characterized by the following sequence: a Guinier regime in the low q range associated with the finite size and mass of the scattered objects, one intermediate regime in which the q dependence is described by a power law with an exponent close to -1, a second Guinier regime at higher q corresponding to the cross-section of the assemblies, and finally well-defined oscillations associated to the shape-dependent form factor of the particle cross-section.

For the data lying in intermediate and low-$q$ Guinier regimes, one can use the Fisher-Burford expression [19] that is well suited to study fractal colloidal aggregates:

$$I(q) = I(0) \frac{1}{(1+\frac{2}{3D_f}q^2 R_G^2)^{D_f/2}} \quad (1)$$

where $D_f$ represents the exponent of the power law in the intermediate regime. Fit of the above expression to the scattering curve is shown in **Figure 2b**, imposing for $D_f$ the value of 1 characteristic of rigid rod. At low $qR_G$ equation 1 is equivalent to the well-known Guinier expression and the fit provides a value of the radius of gyration, $R_G$, and of the zero-wave vector scattered intensity, $I(0)$, in good agreement with the initial slope of the curve $I^{-1}=f(q^2)$ shown in the inset. We obtain $R_G$=77±8 nm, and $M_W$=(31±3)×10$^6$ g/mol, the different values within the uncertainty depending on the method.

Knowing now $D_f$, we propose to model the scattering of the NPs self-assemblies over the whole $q$ range by the form factor of fractal objects with the following expression [20]:

$$\frac{I(q)}{(\Delta\rho)^2 \phi} = VP(q) = N_{agg} q^{-D_f} \frac{\int_0^\infty P_{Spheres}(q,R)L(R,\sigma)R^3 dR}{\int_0^\infty L(R,\sigma)R^3 dR} \quad (2)$$

where $N_{agg}$ is the number of silica nanoparticles inside the complexes, $D_f$ is the fractal dimension of the aggregates, and $P(q)_{spheres}$ is the sphere form factor (see eq 7 in the Experimental Section). We assume the structure factor between aggregates to be close to 1. From the fit of Figure 2a (upper curve), we obtain $R=9.2$ nm, $\sigma=0.12$, $D_f=1$, and $N_{agg}=10$. These results call for two conclusions. First, SiNPs self-assemble in a well-ordered 1D geometry. Secondly, the overall nanoobjects are single-strands with no lateral associations, as shown by the cross-section radius and polydispersity values that are similar to those of free SiNPs.

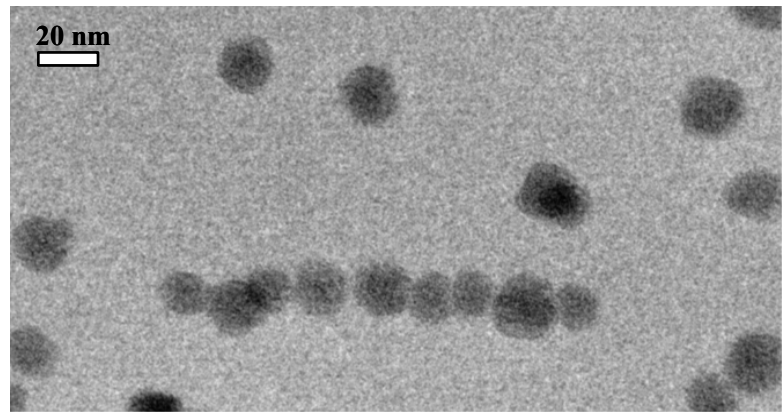

Figure 3. Cryo-TEM image of $C_{chitosan}=0.01$g/l / $C_{SiNP}=10$g/l solutions with $C_{CH3COOH} = 0.3$ M and $C_{CH3COONa} = 0.2$ M.

Cryo-transmission electron microscopy (cryo-TEM) was used to confirm the detailed structure of the soluble complexes in the direct space. **Figure 3** shows a representative TEM image of the chitosan-SiNPs complexes surrounded by individual nanoparticles. It appears that all complexes consist of 1D array of SiNPs in contact with each other inside the complexes. According to the analysis of a number images, the nanometric rods are single strands with average characteristics such as: $<N_{agg,Cryo}> \sim 9$ and $<L_{agg,Cryo}> = 166$ nm. Since the rod radial radius appears constant in the pictures, and branched structures are not observed, we conclude that nanorods correspond to single strands of NPs, rather rigid. These observations agrees with SAXS analysis giving $L_{rod}=2R \times N_{agg}=184$ nm, as an average over the whole sample, in situ in the solution.

Unfortunately the low electronic contrast between the biopolymer and the surrounding medium, grid membrane for TEM or water for SAXS, does not enable us to clearly identify the number of chain per complex and their arrangement in the nanorods structure. However SAXS sheds some light on this aspect, by allowing us to estimate the amount per complexes of silica, and -more indirectly- of chitosan. A simple calculation has been made on the basis of the average number of individual nanoparticles and of nanorods determined by fitting the scattering pattern by the following equation:

$$I(q) = \phi_{rods}I_{rods}(q) + \phi_{indivNPs}I_{indivNPs}(q) \qquad (3)$$

where $I_{rods}(q)$ and $I_{indivNPs}(q)$ are the scattered intensities related respectively to the form factor of the rods and of the individual spherical SiNPs. The best agreement with the data of Figure 2a is obtained for $\phi_{indivNPs}=4.3\times10^{-3}$, and $\phi_{rods}=2.3\times10^{-4}$, and for a number of silica particles inside the nanorods slightly larger, $N_{agg}=14$. Then, considering that SiNPs are in excess and that, according to cryo-TEM, only SiNPs present in complexes display the ribbed texture attributed to chitosan chain binding, one can hypothesize that all the chitosan chains are involved in the complexes. Finally, one derives an average concentration of 1.8 chains of chitosan per nanorods. This result first shows that very few chains are needed to allow SiNPs organisation into nanorods. A second striking point is that the contour length of the biopolymers (~943 nm in weight average) is significantly higher than the average nanorods length (~184 nm) suggesting that chitosan chains are somehow wrapped around SiNPs. This wrapping could occur owing to the adequation between the chitosan chains flexibility and the NPs surface curvature. This is driven by the persistence length, and a simple calculation shows that approximately 8-10 persistence lengths ($L_p$~7.5 nm) are necessary to make a complete turn around a NP ($2\pi R$), suggesting that the chain wrapping around the 10-14 SiNPs into a nanorod is somehow helical ($10 \times L_p \times N_{agg} \approx 2\pi R \times N_{agg} \approx L_{chitosan}$). This explains the single strand rodlike structure.

In conclusion, we have demonstrated by combining SAXS and cryo-TEM that well defined nanorods complexes, quasi-monodisperse in radius, can be obtained by simply mixing cationic chains of natural chitosan with an excess of anionic SiNPs. These 1D aggregates are stable, single-strand and composed in average of 10 SiNPs for 1.8 chitosan chains. The mechanism of these nanorods formation is striking since, to the best of our knowledge, this is the first time that such kind of nanorods are prepared with a semi-flexible polyelectrolyte displaying a persistence length as low as ~8 nm. We point that the ratio between chitosan persistance length and SiNPs radius, $L_p/R$, which is here close to one, may be a determining condition to obtain such objects. Experiments are under course to clarify the global mechanism of these nanorods formation. In the future, we hope that the identification of this parameter $L_p/R$ will open the way toward the control clustering of individual nanoparticles into complexes with a tailored shape.

**Experimental Section**

*Small-Angle X-Ray Scattering Experiments (SAXS):* The SAXS experiments were performed at the ESRF (Grenoble, France) on the ID-02 instrument using the pinhole camera at the energy of 12.46 keV at two sample-to-detector distances (1m and 8m) corresponding to a q-range varying between 0.0011 Å$^{-1}$ and 0.57 Å$^{-1}$. The absolute units are obtained by normalization with respect to water (high *q*-range) or lupolen (low q-range) standard. The total scattered intensity, *I(q)*, of colloidal objects – neglecting the chitosan chains signal- can be expressed by the following equation:

$$I(q) = \phi(\Delta\rho)^2 V P(q) S(q) \qquad (4)$$

where $q = 4\pi/\lambda \times \sin\theta/2$ is the wave vector, $\phi$ is the volume fraction, $(\Delta\rho)^2 = (\rho - \rho_{solvent})^2$ the contrast term, *V* the volume of the scattered objects (related to the weight-average molecular weight M$_W$ of the objects), *P(q)* the form factor and *S(q)* the structure factor. In the first approximation we will consider that inter-object interactions are negligible (diluted regime) and that cross-terms and virial effects are neglected in our fitting procedure (analysis realized in a q-range where $I(q) \sim P(q)$). For SAXS, the scattering length densities (SLDs) are defined by $\rho = 1/(mv \times 1.66 \times 10^{-24}) \times r_{el} \times \Sigma n_i Z_i$, where $r_{el} = 0.28 \times 10^{-5}$ nm is the electron radius, $Z_i$ the atomic number of element i, *m* the monomer mass and *v* the monomer specific volume (0.478 cm$^3$/g for chitosan and 0.4545 cm$^3$/g for silica). For chitosan and SiNPs we found $\rho_{chitosan} = 18.7 \times 10^{10}$ cm$^{-2}$ and $\rho_{SiNP} = 18.5 \times 10^{10}$ cm$^{-2}$, respectively.

We first characterized the scattering from a SiNPs suspension, introducing a polydispersity in size of the scattered objects described by a log-normal distribution, $L(r, R, \sigma)$, where *r* is the radius, *R* the mean radius, and $\sigma$ the variance:

$$L(r, R, \sigma) = \frac{1}{\sqrt{2\pi} r \sigma} \exp\left(-\frac{1}{2\sigma^2} \ln^2\left(\frac{r}{R}\right)\right) \qquad (5)$$

Thus, neglecting the virial effects (assuming $S(q) = 1$) at low concentration in presence of salt, it is classical to define the global scattering intensity by the following relation:

$$I(q) = \phi(\Delta\rho)^2 V \int_0^\infty P(q, r) L(r, R, \sigma) dr \qquad (6)$$

Figure 2a shows the scattering of the pure SiNPs solution (lower curve), which can be fitted satisfactorily by means of the form factor expression derived for hard spheres of radius $R$:

$$P(q) = 9 \times \left[ \frac{\sin(qR) - qR\cos(qR)}{(qR)^3} \right]^2 \qquad (7)$$

The form factor oscillations, damped by the size distribution, are well reproduced with $I(q)$ calculated as indicated above (eqs 6 and 7). The SiNPs solution is well represented by a suspension of hard spheres with $R$=9.2 nm, and σ=0.12. Extrapolation of the scattered intensity to zero-wave vector, $I(0)$, gives the weight-average molecular weight, $M_{W,SiNP}$=3×10$^6$ g/mol.

The discussion of the signal from the complexes is given in the text.

*Cryogenic Transmission Electron Microscopy:* Cryo-transmission electron microscopy (cryo-TEM) was performed on vitrified complexes prepared at r = 0.01. In brief, a drop of the solution to be imaged was poured onto a TEM carbon grid covered by a 100 nm thick polymer perforated membrane. The drop was blotted with filter paper, and the grid was quenched rapidly in liquid ethane in order to avoid the crystallization of the aqueous phase. The vitrified samples were then stored under liquid nitrogen and transferred to the vacuum column of a Tecnai TEM microscope operating at 120 kV. The magnification for the cryo-TEM experiments was selected at 40 000 × .

**Table of contents:** We show by combining SAXS and cryo-TEM that anionic silica nanoparticles (SiNPs) assemble into well-defined 1D cluster when mixed with a dilute solution of semi-flexible chitosan polycation. The nanorods are stable in excess of SiNPs and composed of 10 SiNPs well ordered into straight single strands with length $L_{rod} \approx 184.0$ nm and radius $R_{rod} = 9.2$ nm $= R_{SiNPs}$. We point that the ratio between chitosan persistence length and SiNPs radius, which is here equal to one, can be the determining condition to obtainsuch original objects.

Keywords: Nanoparticles, Biopolymers, Complexation, Self-assemblies, Small-angle X-ray scattering

Li Shi,[1,2] Florent Carn,[1,*] François Boué,[2] Gervaise Mosser,[3] and Eric Buhler[1,2,*]

Nanorods of Well-Defined Length and Monodisperse Cross-Section Obtained from Electrostatic Complexation of Nanoparticles with a Semiflexible Biopolymer

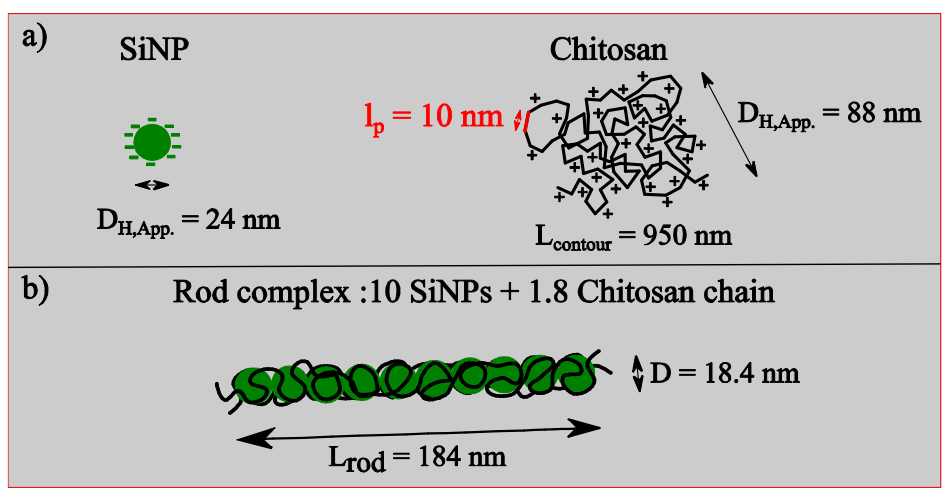